# Strongly linked current flow in polycrystalline forms of the new superconductor $MgB_2$

**D.C. Larbalestier*[†], M. O. Rikel*, L.D. Cooley*, A.A. Polyanskii*, J.Y. Jiang*, S. Patnaik*, X.Y. Cai*, D.M. Feldmann*, A. Gurevich*, A.A. Squitieri*, M.T. Naus*, C. B. Eom*[†] E.E. Hellstrom,*[†]**

** Applied Superconductivity Center, University of Wisconsin–Madison, 1500 Engineering Drive, Madison, WI 53706*

*† Department of Materials Science and Engineering, University of Wisconsin–Madison, Madison, WI 53706*

**R.J. Cava, K.A. Regan, N. Rogado, M.A. Hayward, T. He, J.S. Slusky, P. Khalifah, K. Inumaru, and M. Haas**

*‡ Department of Chemistry and Princeton Materials Institute, Princeton University, Princeton NJ 08544*

**The discovery of superconductivity at 39 K in $MgB_2$[1] raises many issues. One of the central questions is whether this new superconductor resembles a high-temperature-cuprate superconductor or a low-temperature metallic superconductor in terms of its current carrying characteristics in applied magnetic fields. In spite of the very high transition temperatures of the cuprate superconductors, their performance in magnetic fields has several drawbacks[2]. Their large anisotropy restricts high bulk current densities to much less than the full magnetic field-temperature (H-T) space over which superconductivity is found. Further, weak coupling across grain boundaries makes transport current densities in untextured polycrystalline forms low and strongly magnetic field sensitive[3,4]. These studies of $MgB_2$ address both issues. In spite of the multi-phase, untextured, nano-scale sub-divided nature of our samples, supercurrents flow throughout without the strong sensitivity to weak magnetic fields characteristic of Josephson-coupled grains[3]. Magnetization measurements over nearly all of the superconducting H-T plane show good temperature scaling of the flux pinning force, suggestive of a current density determined by flux pinning. At least two length scales are suggested by the magnetization and magneto optical (MO) analysis but the cause of this seems to be phase inhomogeneity, porosity, and minority insulating phase such as MgO rather than by weakly coupled grain boundaries. Our results suggest that polycrystalline ceramics of this new class of superconductor will not be compromised by the weak link problems of the high temperature superconductors, a conclusion with enormous significance for applications if higher temperature analogs of this compound can be discovered.**

The principal samples were synthesized by direct reaction of bright Mg flakes (Aldrich Chemical) and sub-micron amorphous B powder (Callery Chemical). Starting materials were lightly mixed in half-gram batches, and pressed into pellets. These pellets were placed on Ta foil, which was in turn placed on $Al_2O_3$ boats, and fired in a tube furnace under a mixed gas of 95% Ar 5% $H_2$ for 1 hour at 600 C, 1 hour at 800 C, and 1 hour at 900 C, and then lightly ground. The resulting powders were pressed into pellets and then hot pressed at 10 kbar at temperatures between 650 and 800 °C for periods between 1 and 5.5 hours. Disks ~4 mm in diameter and ~1 mm thick were cut from these pellets for property characterization. As noted later, this process cannot yet be considered optimum.

Magnetization properties were examined in SQUID and vibrating sample magnetometers (VSM) in applied fields up to 14 T from 4.2 to above $T_c$. Figure 1 shows onset $T_c$ values of 37-38 K for the above samples and for commercial $MgB_2$ powder (99.5%, ~2 μm diameter by examination, CERAC). Sample 1 and the commercial powder show smooth transitions with some temperature dependence of the zero-field cooled (ZFC) shielded moment, while sample 3 exhibits a step, indicative of non-uniformity in superconducting properties. These transitions are not as sharp as those reported by Bud'ko et al.[5] for $MgB_2$ prepared in sealed Ta and quartz ampules.

Figure 2 shows results of the large *H-T* range VSM examination. Large hysteresis characteristic of bulk currents was seen at all temperatures; however the major part of the hysteresis loops closed at fields about half of the upper critical field $H_{c2}(T)$. A smaller hysteretic tail which closes at ~ ¾ $H_{c2}(T)$ can also be seen in Fig. 2 inset. $H_{c2}(T)$ was determined as the field at which the moment first deviated from the background, as indicated by the dashed line. $H_{c2}(T)$ appears to vary more slowly than predicted by the Werthamer-Helfand-Hohenberg model[6] which would predict $\mu_0 H_{c2}(0) = 0.7\mu_0 T_c (dH_{c2}/dT)|_{Tc} \approx 14$ T from the rather modest slope of $\mu_0 H_{c2}(T)$ of ~0.5 $TK^{-1}$ at $T_c$. In fact $\mu_0 H_{c2}(0)$ appears to be 17.5 T, giving a zero temperature coherence length, $\xi(0) = [\phi_0/2\pi\mu_0 H_{c2}(0)]^{0.5}$ of ~4 nm. The value of $\xi(0)$ is thus larger than the 1-2 nm values for typical HTS compounds. These results are consistent with the observations of Finnemore et al.[7] at 30 and 36 K.

The flux-pinning characteristics follow a Kramer-like function similar to that of $Nb_3Sn$[8], where $J_c^{0.5}H^{0.25}$ is linear in $H$ (Fig. 3). Scaling of the flux pinning force, $F_p = J_c\mu_0 H$, is found down to at least 20 K for sample 1 (Fig. 3 inset). The extrapolated Kramer curve intercept $H_K(T)$ defines an empirical irreversibility line for which $J_c$ tends to zero. The $H_K(T)$ line is directly proportional to $\mu_0 H_{c2}(T)$ (Fig. 2), a result very different from the $(T_c - T)^{1.5}$ dependence seen for the irreversibility field $H^*(T)$ in HTS materials[2]. The proportionality of $H^*(T)$ and $\mu_0 H_{c2}(T)$ is similar to that observed for Nb-Ti and $Nb_3Sn$[9]. However, the curvature of the Kramer extrapolation at the highest fields is also suggestive of a weaker superconducting phase that is controlling the flow of currents through the matrix. Even so, the magnetization hysteresis width in both SQUID and VSM curves require whole-sample current densities of order $10^4$ $Acm^{-2}$ at 25 K, 1 T, increasing to ~$4\times10^4$ $Acm^{-2}$ at 4.2 K, 1 T, using standard Bean formulae[10].

To check for the possibility of anisotropy in this system, samples were cut from two orthogonal axes relative to the hot press axes. Magnetization measurements on these samples yielded nearly identical values of $\mu_0 H_{c2}(T)$, suggesting a lack of texture in the samples. No texture was indicated by x-ray study of these same samples.

The issue of inhomogeneous and granular behavior was assessed by correlated magneto-optical (MO) and analytical scanning electron microscopy, supported by direct transport measurements, and analysis of the remanent magnetization in a SQUID magnetometer using the method of Müller et al.[11], as shown in Fig. 4. The remanent-field data were obtained by first cooling to 5 K in zero field and then measuring the moment due to the residual flux pinning for various applied fields. The two characteristic fields for flux penetration, indicated by the arrows, show the distinctively different fields required for flux penetration first into the center of the whole sample and then into the interior of the stronger superconducting regions themselves. Taking the whole sample radius of 2 mm, the bulk penetration field of ~ 20 mT corresponds to a bulk shielding current of $0.7\times10^3$ $Acm^{-2}$ (0 T, 5 K). Similarly, taking the high penetration field value of 0.5 T and the characteristic size of the strongly shielding regions in Fig. 5b as 150 µm, $J_c$ = 0.3 $MAcm^{-2}$ at 0 T, 5 K is indicated.

Comparison of the light and MO images in Figs. 5a and 5b shows a complex multi-scale microstructure. X-ray analysis showed only ~5% MgO as a minority phase. The MO images all show direct evidence of superconducting inhomogeneity, as exemplified in Fig. 5b, where very strong superconducting regions of sizes up to about ~150 µm form a minority fraction on several length scales throughout the whole sample. The flux gradient d$H$/d$x$ across these regions can be observed at all temperatures from 11 to 38 K. The maximum gradients correspond to $10^5$ $A/cm^2$ at 11 K, 0.05 T.

Figs. 5c and 5d investigate the microstructure of the strongest superconducting regions and show that they are multiphase, too. In fact these strong superconducting regions are subdivided on a scale of ~100 nm, perhaps due to partial decomposition in situ. Electron microprobe analysis verifies that the darker central area corresponds to a mixture of $MgB_2$ and boron-rich phases. Since none of the x-ray examinations suggest any texture, we conclude that these regions consist of an untextured two-phase nanomixture of $MgB_2$, MgO, pores and perhaps some amorphous B-rich phase. The important implication is that there is a very large number of high-angle grain boundaries and blocking insulating phases within each of the strongly shielding regions. Nonetheless, these regions support high current densities of order $10^5$ $Acm^{-2}$, which they could not sustain if there was any inherent strong suppression of current across the grain boundaries. Indeed, the true local current densities within the superconducting particles must be significantly larger.

The totality of our data thus leads to the conclusion that $MgB_2$ is more akin to a low-$Tc$ metallic than to a high-$T_c$ cuprate superconductor. A far-from-single-phase sample nevertheless has large current densities that circulate over lengths that are many grain sizes, whether we consider the ~100 nm grains within the strongest ~100 µm agglomerates, or the scale of the whole sample. The basic $H$-$T$ boundary of the superconducting phase has been mapped, with $H_{c2}(0)$ attaining 17.5 T. This value exceeds that of Nb-Ti but not that of $Nb_3Sn$ and is rather low for a material with $T_c$ of 38 K. However, the combination of much higher $T_c$ than Nb-base materials and the apparent lack of granularity means that new compounds based on this system could have interesting values of $T_c$, $H_{c2}$, and most important of all $J_c$ without the need for the high degree of texture that dominates all polycrystalline use of high-$T_c$ materials.

This work has been supported by the Air Force Office of Scientific Research, the Department of Energy, and the National Science Foundation through the MRSEC program.

[1] J. Akimitsu, Symposium on Transition Metal Oxides, Sendai, January 10, 2001, J. Nagamatsu, N. Nakagawa, T. Muranaka, Y. Zenitani, and J. Akimitsu (submitted).

[2] D.S. Fisher, M.P.A. Fisher, and D.A. Huse, “Thermal fluctuations, quenched disorder, phase transitions, and transport in type-II superconductors”, Phys. Rev. B 43, 130-159 (1991).

[3] D. Dimos, P. Chaudhari, and J. Mannhart, “Superconducting transport properties of grain boundaries in $YBa_2Cu_3O_7$ bicrystals”, Phys. Rev. B 41, 4038-4049 (1990).

[4] N. F. Heinig, R. D. Redwing, J. E. Nordman, and D. C. Larbalestier, Phys. Rev B 60, 1409 (1999).

[5] S.L. Bud’ko, G. Lapertot, C. Petrovic, C.E. Cunningham, N. Anderson, and P.C. Canfield, “Boron isotope effect in superconducting $MgB_2$”, Phys. Rev. Lett. (submitted).

[6] N.R. Werthamer, E. Helfand, and P.C. Hohenberg, “Temperature and purity dependence of the superconductign critical field, $H_{c2}$,” Phys. Rev. 147, 288-294 (1966).

[7] D. K. Finnemore, J. E. Ostenson, S. L. Bud’ko, G. Lapertot, P. C. Canfield, cond-mat/0102114

[8] D. Dew-Hughes, Phil. Mag. **55**, 459-479 (1987).

[9] M. Suenaga, A.K. Ghosh, Y. Xu, and D.O. Welch, “Irreversibility temperatures of Nb3Sn and Nb-Ti”, Phys. Rev. Lett. 66, 1777-1780 (1991)

[10] C.P. Bean, Rev. Mod. Phys. 36, 31 (1964).

[11] K–H. Muller, C. Andrikidis, J. Du, K. E. Leslie, and C. P. Foley, Phys. Rev. B 60, 659 (1999)

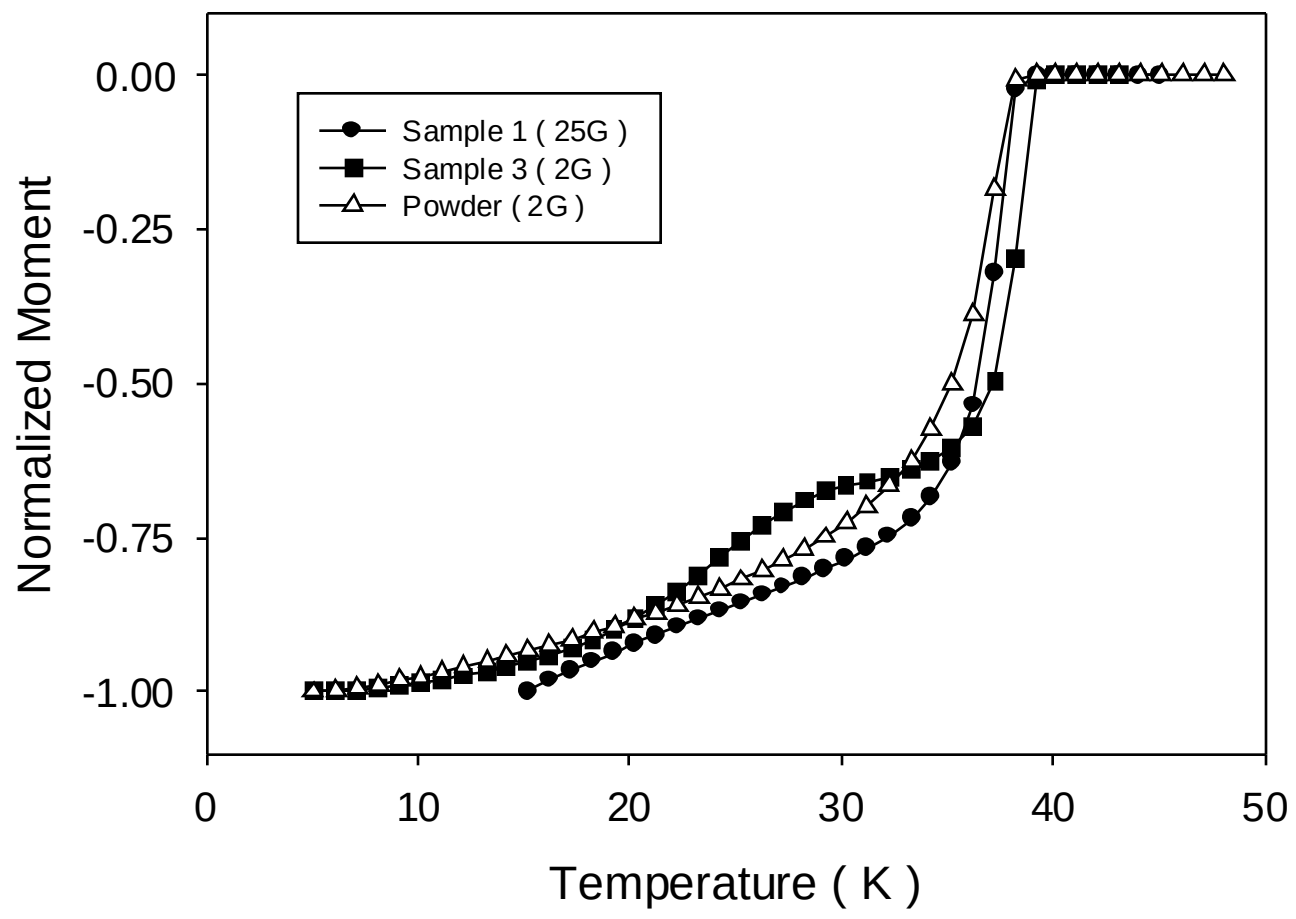

Fig. 1. ZFC magnetization (normalized to full screening at 4 K) for the different $MgB_2$ samples studied. The measuring field is indicated in the key. All samples exhibit onset Tc of 37-38 K and broadened transitions indicative of sample inhomogeneity.

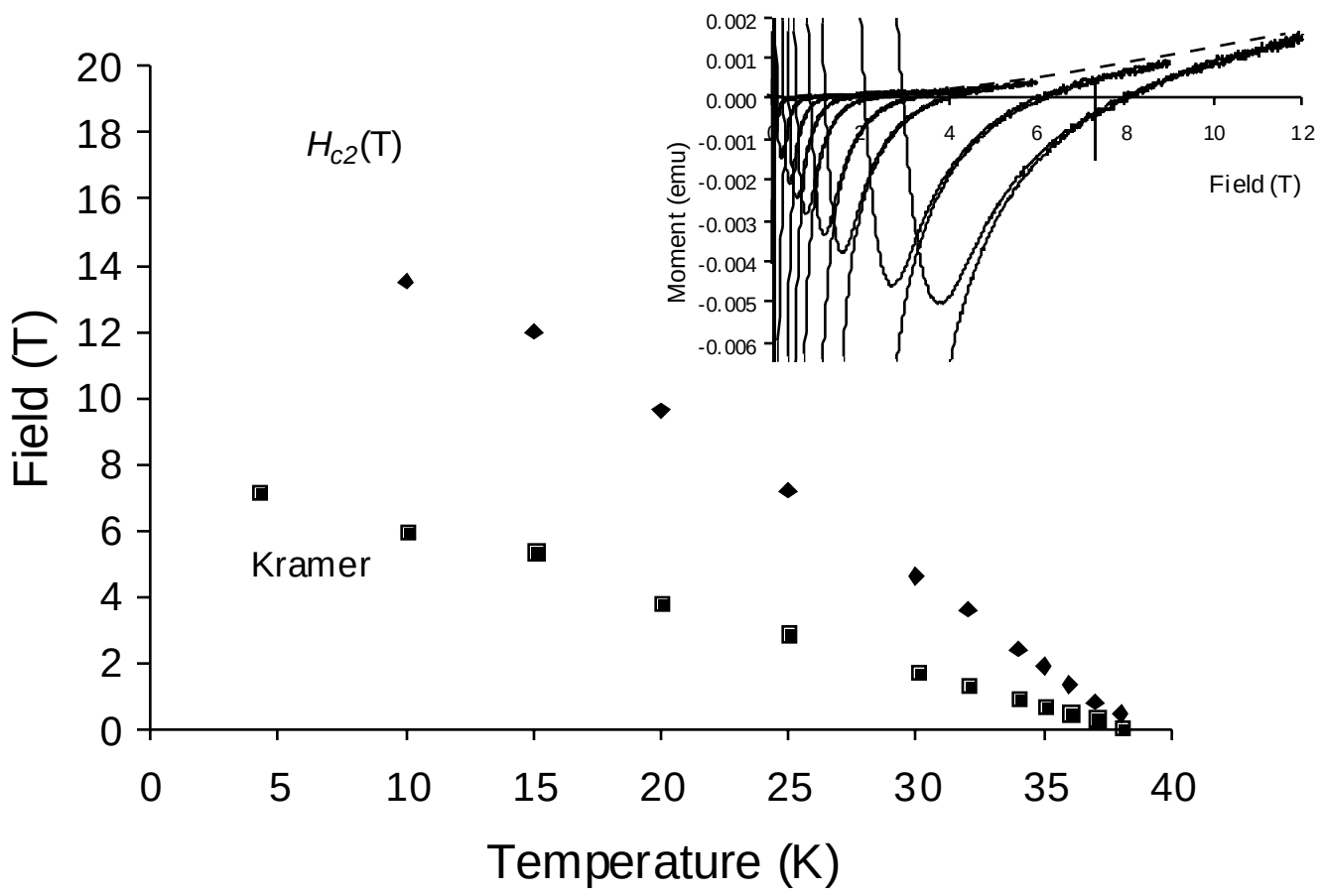

Fig. 2: Upper critical field (◆) and Kramer extrapolation field (■) are plotted as a function of temperature. Inset: VSM data for sample 1 at 38, 37, 36, 35, 34, 32, 30, 25, and 20 K. The dashed line represents the background above $H_{c2}(T)$.

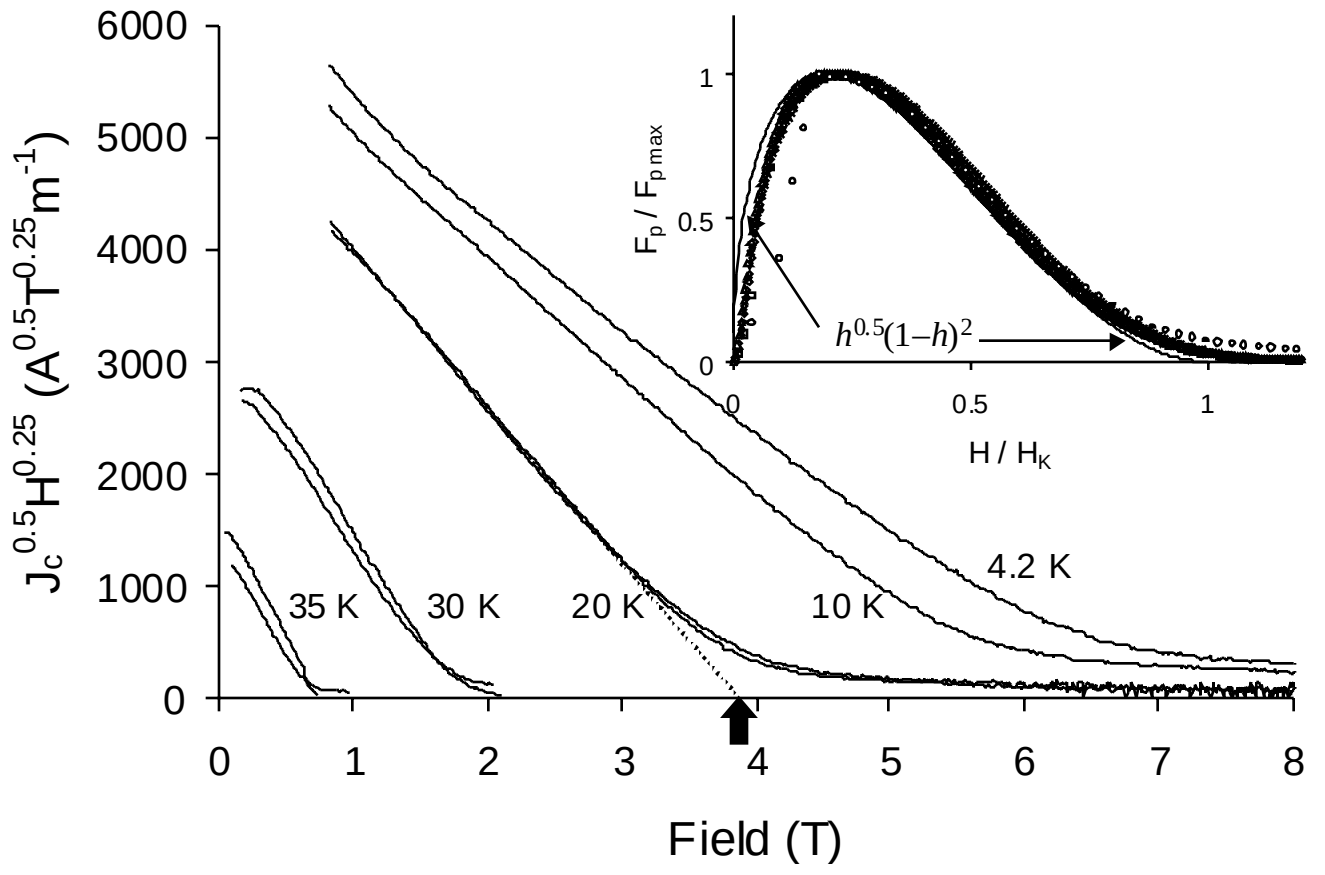

Fig. 3: Kramer curves for samples 1 (light lines) and 3 (heavy lines) at 35 to 4.2 K. The dashed line and the arrow indicate the extrapolated value $H_K$ at 20 K. Inset: Scaling of the reduced bulk pinning force data for sample 1 at 37 (O), 35 (Δ), 30 (+), 25 (×), and 20 K (◆).

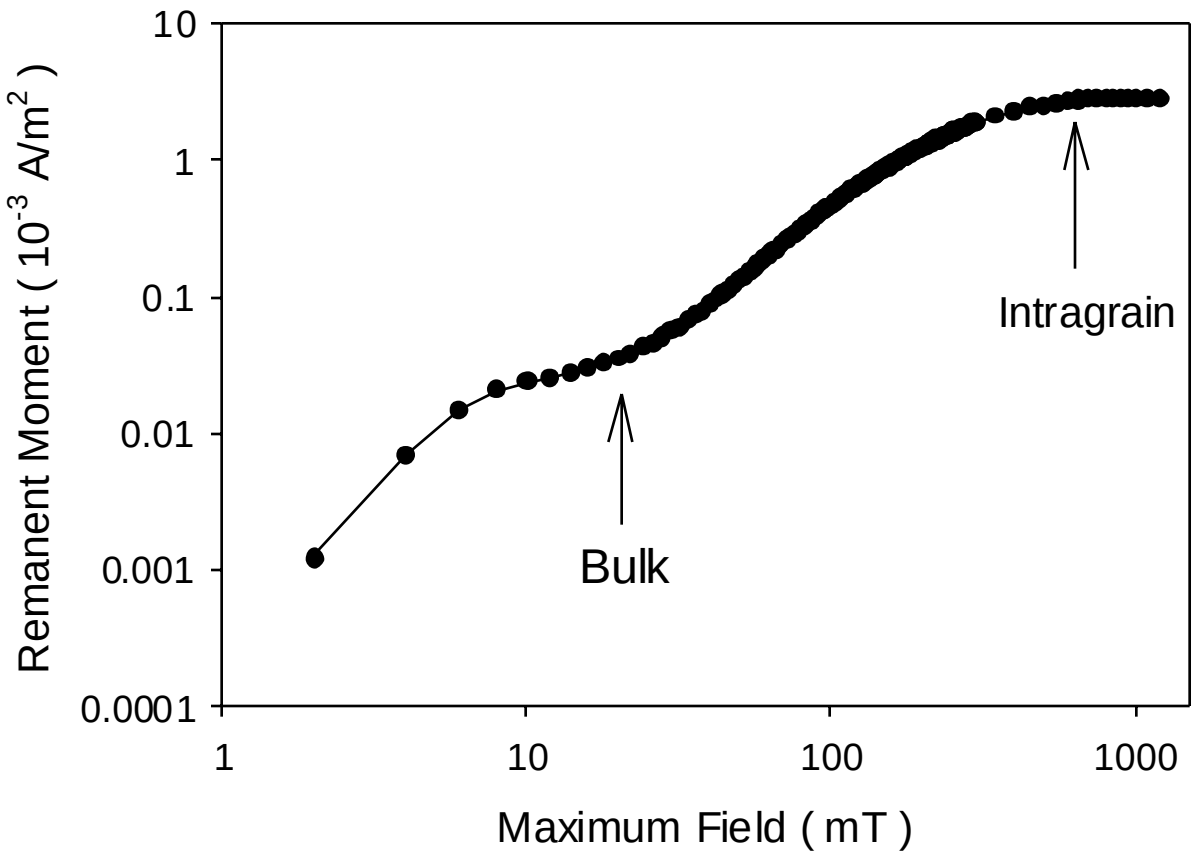

Fig. 4. Total remanent moment due to screening of a ZFC applied field at 5 K for sample 1. The arrows indicate flux penetration into the sample center and into the center of the strongly superconducting regions, respectively.

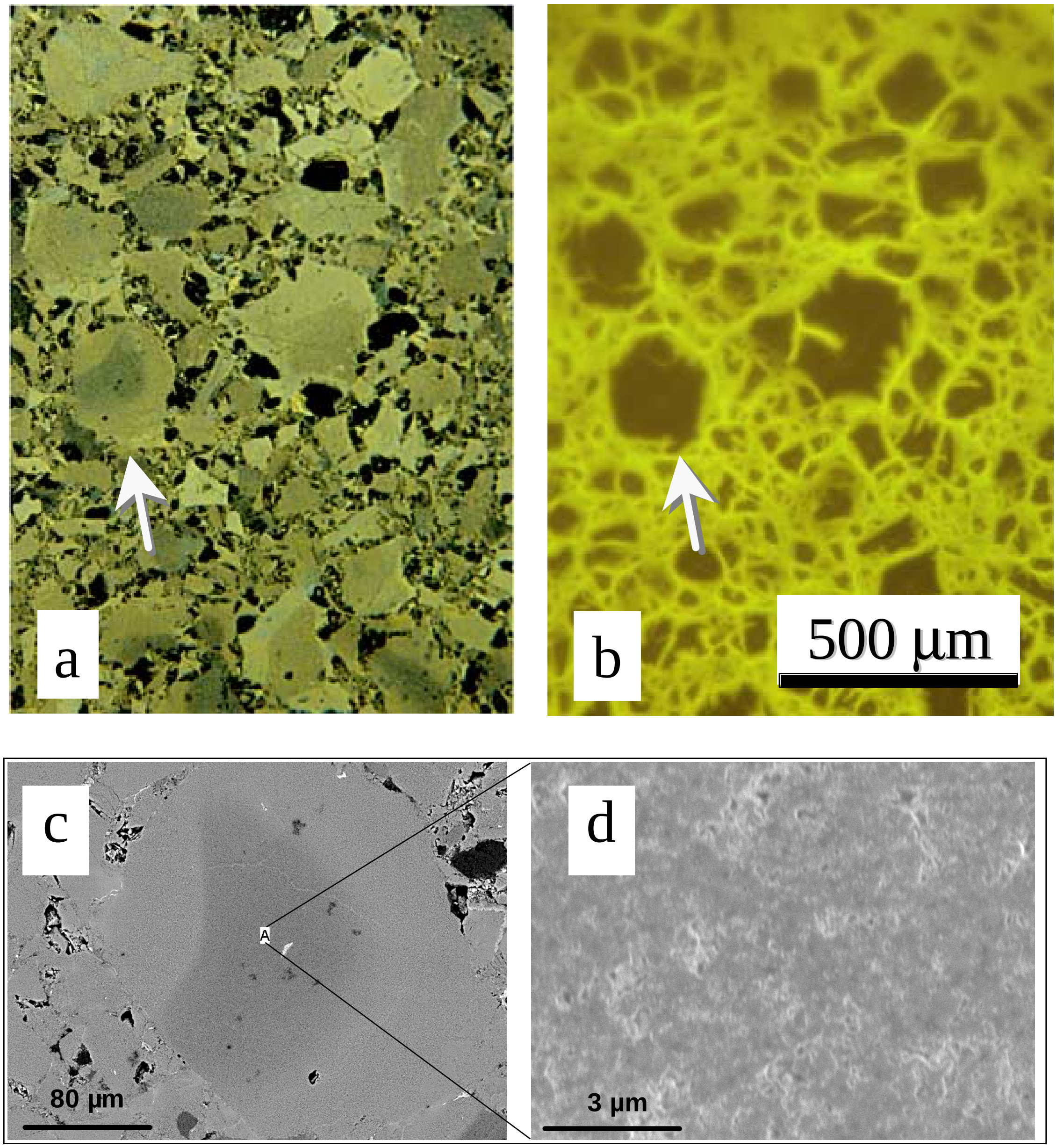

Fig. 5: (a) Light microscopy and (b) magneto-optical of the same area of sample 3 showing extensive inhomogeneity of the sample. In (b), the dark regions exhibit very strong superconductivity, the lighter areas weaker superconductivity. (c) SEM backscattered electron image of a strongly superconducting region marked with the arrrow in (a) and (b). (d) Higher resolution secondary electron SEM examination of the region in (c) reveals that the area marked by the arrow in (a) and (b) has fine scale structure.